\documentclass[a4paper]{article}

\usepackage{INTERSPEECH2021}
\usepackage{enumerate}
\usepackage{breqn}
\usepackage{algorithm}
\usepackage[noend]{algpseudocode}
\usepackage{multirow}
\usepackage{soul,color}
\usepackage{balance}

\title{Towards Robust Speech-to-Text Adversarial Attack}
\name{Mohammad Esmaeilpour, Patrick Cardinal, Alessandro Lameiras Koerich}
\address{
  \'{E}cole de Technologie Sup\'{e}rieure (\'{E}TS),
D\'{e}partement de G\'{e}nie Logiciel et des TI\\
1100 Notre-Dame West, Montr\'{e}al, H3C 1K3, Qu\'{e}bec, Canada
  }
\email{mohammad.esmaeilpour.1@ens.etsmtl.ca,\{patrick.cardinal, alessandro.koerich\}@etsmtl.ca}

\begin{document}

\maketitle
\begin{abstract}
This paper introduces a novel adversarial algorithm for attacking the state-of-the-art speech-to-text systems, namely DeepSpeech, Kaldi, and Lingvo. Our approach is based on developing an extension for the conventional distortion condition of the adversarial optimization formulation using the Cram\'{e}r integral probability metric. Minimizing over this metric, which measures the discrepancies between original and adversarial samples' distributions, contributes to crafting signals very close to the subspace of legitimate speech recordings. This helps to yield more robust adversarial signals against playback over-the-air without employing neither costly expectation over transformation operations nor static room impulse response simulations. Our approach outperforms other targeted and non-targeted algorithms in terms of word error rate and sentence-level-accuracy with competitive performance on the crafted adversarial signals' quality. Compared to seven other strong white and black-box adversarial attacks, our proposed approach is considerably more resilient against multiple consecutive playbacks over-the-air, corroborating its higher robustness in noisy environments.

\end{abstract}

\section{Introduction}
During the last years and especially after the characterization of adversarial attacks for the computer vision applications \cite{szegedy2013intriguing}, several investigations have been conducted on generalizing this threat to the audio recognition and speech transcription models \cite{esmaeilpour2020detection,carlini2018audio,qin2019imperceptible,esmaeilpour2019robust}. It has been proven that adversarial signals exist for both 1D and representation (spectrogram) levels, which can seriously debase the performance of the cutting-edge speech-to-text models such as DeepSpeech \cite{MozillaImplementation}, Kaldi\cite{povey2011kaldi}, and Lingvo \cite{shen2019lingvo}. However, developing effective adversarial signals resilient to environmental noises and room settings is challenging \cite{yakura2018robust,szurley2019perceptual}. These settings include the position and characteristics of both the microphone and speaker and the room's geometry. Under various settings, simply playing the crafted adversarial signal over-the-air and recording it by another microphone most likely removes the obtained adversarial perturbation \cite{carlini2018audio}. For addressing this issue, several expectation over transformation (EOT) operations have been introduced \cite{qin2019imperceptible,schonherr2020imperio,chen2020metamorph,abdullah2019practical}. These operations often employ room filter sets (e.g., channel impulse response \cite{chen2020metamorph}) as part of the adversarial optimization procedure to avoid bypassing the perturbation after playing over-the-air. However, developing EOT is dependent on some static room assumptions, which might negatively affect the generalizability of the filter sets \cite{schonherr2020imperio,esmaeilpour2020class}. 

In a big picture, the optimization formulation toward crafting an adversarial signal for a speech-to-text model has two parts: (i) optimization term and (ii) the distortion condition (relative constraint), as follows~\cite{carlini2018audio}:
\begin{equation}
    \underbrace{\min_{\delta} \left \| \delta \right \|_{F} + \sum_{i}c_{i}\mathcal{L}_{i}(\vec{x}_{\mathrm{adv}},\hat{\mathbf{y}}_{i})}_{\text{optimization term}} \quad \mathrm{s.t.} \quad \underbrace{l_{\text{dB}}(\vec{x}_{\mathrm{adv}})}_{\text{distortion condition}} < \epsilon
    \label{eq:generalformulation}
\end{equation}
\noindent where $\delta$ is the adversarial perturbation achievable through this iterative procedure for the original input signal $\vec{x}_{\mathrm{org}}$ to yield the adversarial signal $\vec{x}_{\mathrm{adv}}$ ($\vec{x}_{\mathrm{adv}} = \vec{x}_{\mathrm{org}} + \delta$). Additionally, $c_{i}$, $\epsilon$, and $\hat{\mathbf{y}}_{i}$ are the scaling coefficient, audible threshold, and the targeted incorrect phrase defined by the adversary, respectively. Furthermore, $\mathcal{L}(\cdot)$ denotes the loss function such as the connectionist temporal classification (CTC) loss \cite{graves2006connectionist,carlini2018audio}, the psychoacoustic loss function \cite{szurley2019perceptual}, cross entropy loss \cite{qin2019imperceptible}, etc. In this typical formulation, the distortion condition is usually known as the loudness metric $l_{\text{dB}}(\cdot)$ computed in the logarithmic $\mathrm{dB}$-scale with respect to the human hearing range \cite{carlini2018audio}. 

The EOT operations incorporated in the state-of-the-art adversarial attack algorithms often involve the optimization term in Eq.~\ref{eq:generalformulation} \cite{qin2019imperceptible,schonherr2020imperio,chen2020metamorph}. Herein, we discuss extending the distortion condition in this equation to avoid implementing the costly EOT-based operations applied on the optimization term. This also helps to craft more robust adversarial signals. Toward this end, we review some strong adversarial attack approaches in Section~\ref{sec:background}. Then, we provide theoretical explanations on developing a relative constraint (the distortion condition) in Section~\ref{sec:proposed}. Finally, we analyze the achieved results from the conducted experiments on attacking speech-to-text models in Section~\ref{sec:experiments}. In summary, we make the following contributions in this paper:
\begin{enumerate}[(i)]
    \item developing an extension for the distortion condition of an adversarial attack formulation using the Cram\'{e}r integral probability metric;
    \item introducing a white-box attack framework for crafting adversarial signal more robust against over-the-air playbacks; 
    \item avoiding time-consuming room impulse response simulations and costly EOT operations in the adversarial optimization formulation (i.e., Eq.~\ref{eq:generalformulation}).
\end{enumerate}

\section{Background}  
\label{sec:background}
In this section, we review the cutting-edge white and black-box adversarial attack algorithms developed against speech-to-text models. More specifically, we focus on the EOT-based attacks since they are, to some extent, capable algorithms in crafting over-the-air resilient adversarial signals \cite{qin2019imperceptible}. However, we start with the baseline EOT-free C\&W attack \cite{carlini2018audio} developed for the DeepSpeech speech-to-text system. This algorithm is based on Eq.\ref{eq:generalformulation} and introduces a simple yet effective distortion condition for a targeted attack scenario as the following \cite{carlini2018audio}.
\begin{equation}
l_{\text{dB}}(\vec{x}_{\mathrm{adv}})=l_{\text{dB}}(\delta)-l_{\text{dB}}(\vec{x}_{\mathrm{org}})  
\label{eq:distortionSimple}
\end{equation}
\noindent where $l_{\text{dB}}(\cdot)$ can be scaled by factor of 20 to better fit the human audible range \cite{carlini2018audio}. The C\&W attack uses the CTC loss function with an assumption of optimizing $\min_{\delta} \left \| \delta \right \|_{2}^{2}$ for the string tokens $\pi_{i}$ (without duplication) which eventually should reduce to $\hat{\mathbf{y}}_{i}$ (after greedy or beam search decoding \cite{carlini2018audio}). Although this distortion metric constraints the C\&W algorithm to craft an adversarial signal almost seamless to the original sample $\vec{x}_{\mathrm{org}}$, it does not impose a strict condition to generate an over-the-air resilient adversarial signal. Presumably, this is due to making a reasonable trade-off between adversarial signal quality and attaining small magnitude for the adversarial perturbation $\delta$.

The EOT operation introduced by Qin {\it et al.}~\cite{qin2019imperceptible} uses the acoustic room simulator followed by speech reverberation filtrations for crafting resilient adversarial signals in adverse scenarios (i.e., multiple over-the-air playbacks). This algorithm is known as the Robust Attack, and it fits in the targeted adversarial category incorporating a variety of room settings for improving its performance. The optimization procedure of this attack is as follows \cite{qin2019imperceptible}.
\begin{equation}
    \min_{\delta}\mathbb{E}_{t\sim \tau}\left [ \ell_{net}\left ( \mathbf{y}_{i},\hat{\mathbf{y}}_{i} \right )+c_{i} \ell_{m}(\vec{x}_{org,i},\delta_{i}) \right ] \quad \mathrm{s.t.} \quad \left \| \delta \right \|< \epsilon
    \label{eq:robustAttack}
\end{equation}
\noindent where $\tau$ is the EOT filter set predefined (computed according to the room setting) by the adversary and $\mathbf{y}_{i}\neq \hat{\mathbf{y}}_{i}$ where the latter refers to the ground truth phrase associated with $\vec{x}_{\mathrm{org}}$. Moreover, $\ell_{net} (\cdot)$ and $\ell_{m} (\cdot)$ denote the cross entropy and the masking threshold loss functions, respectively. The Robust Attack has been tested on the Lingvo speech-to-text system and it has demonstrated a high capacity for crafting resilient over-the-air adversarial signals.

Yakura {\it et al.}~\cite{yakura2018robust} introduced a similar EOT operation, which employs band-pass filtration according to the human cut-off hearing range on top of the simulated room impulse response (RIR) filter set. Moreover, this attack implements the white Gaussian noise (WGN) filtration so as to effectively simulate the environmental noises as the following.
\begin{dmath}
    {\min_{\delta} \mathbb{E}_{t\in \tau, \omega\sim \mathcal{N}(0,\sigma^{2})}\left [ \mathcal{L} ( \mathrm{mfcc}(\vec{x}_{\mathrm{adv}},\hat{\mathbf{y}}_{i})+\alpha_{k}\left \| \delta \right \| \right ]}\\ \vec{x}_{\mathrm{adv}}= \left [ \vec{x}_{\mathrm{org}}+\Omega(\delta) \right ]\circledast t+\omega \quad  \mathrm{s.t.} \quad \left \| \delta \right \|< \epsilon
    \label{eq:Yakura}
\end{dmath}
\noindent where $\omega$, $\mathrm{mfcc}$, and $\alpha_{k}$ denote the WGN filter drawn from the normal distribution with variance $\sigma^{2}$, the Mel-frequency cepstral coefficient transform \cite{davis1980comparison}, and the scaling hyperparameter defined by the adversary, respectively. Additionally, $\Omega(\cdot)\in[1,4]$ kHz refers to the band-pass filtration operation and $\circledast$ is the convolution operator. Herein, $\mathcal{L}(\cdot)$ stands for the CTC loss function, and it has been adapted to the DeepSpeech victim model. The reported experiments demonstrated that Yakura's attack outperforms the C\&W in a variety of environmental scenes \cite{yakura2018robust}. However, at the cost of higher computational complexity for computing $\tau$ filter set. 

One reliable approach, which implements the RIR simulation with a relatively lower computational cost is the Imperio attack \cite{schonherr2020imperio}. This algorithm employs a deep neural network (DNN) to simulate RIR filter set and the psychoacoustic thresholding ($pst$) for crafting over-the-air resilient adversarial signals (see Eq.~\ref{eq:imperio} \cite{schonherr2020imperio}).
\begin{equation}
    \vec{x}_{\mathrm{adv}}  = \underbrace{\arg \max_{\vec{x}_{i}}\mathbb{E}_{t\sim \tau_{d}}\left [ P(\hat{\mathbf{y}}_{i}|\vec{x}_{i,t}) \right ]}_{\vec{x}_{\mathrm{org}}+\kappa \left [ \partial \ell_{net}(\mathbf{y},\hat{\mathbf{y}})/ \partial f^{*}(\vec{x}_{\mathrm{org}}) \right ] }
    \label{eq:imperio}
\end{equation}
\noindent where $d$, $\kappa$, and $f^{*}(\cdot)$ denote the dimension of the filter set, the learning rate and the post-activation function of the DNN model mentioned above, respectively. The EOT operation incorporated in the Imperio attack is dynamic and fits well for various room settings including meeting, lecture, and office. The distortion condition in this attack is $\delta \leq pst$ and should be tuned for every incorrect phrase $\hat{\mathbf{y}}$. Imperio has been tested on the Kaldi system. Such an attack has considerably reduced this advanced speech-to-text model's performance even after playback over-the-air.

Since the robustness of an adversarial signal over-the-air can also depend on the characteristics of both speaker and microphone, channel impulse response (CIR) filter set is developed as part of the EOT operation in the Metamorph adversarial attack \cite{chen2020metamorph}. The general formulation of this attack is as the following.
\begin{equation}
    \min_{\delta} \alpha_{t} l_{\text{dB}}(\vec{x}_{\mathrm{adv}})+\frac{1}{M}\mathcal{L}(\vec{x}_{\mathrm{org}}+\delta_{i},\pi_{i}) \quad \mathrm{s.t.} \quad \left \| \delta \right \|< \epsilon
    \label{eq:metamorph}
\end{equation}
\noindent where $\alpha_{t}$ is the balancing coefficient between the quality of the crafted adversarial signal and the overall success rate of the attack algorithm on the victim model. Additionally, $M$ indicates the number of microphone-speaker positions in an enclosed environment. These hyperparameters have a key role in crafting robust adversarial signals, which the adversary should precisely locate. The effectiveness of the Metamorph adversarial attack has been proven for the DeepSpeech system. However, at the cost of employing various CIR filer sets \cite{chen2020metamorph}. 

Developing EOT operations for the black-box adversarial attack is extremely challenging since the adversary does not have access to the victim model and its associated settings. In response to this limitation, an over-the-line technique has been developed to surrogate the over-the-air EOT operations \cite{abdullah2019practical}. However, this technique requires numerous experiments to capture local and global environmental scene distributions. Regarding this concern, there are two EOT-free black-box adversarial attacks with competitive performance to the over-the-line approach in attacking the DeepSpeech system: (i) the genetic algorithm attack (GAA) \cite{taori2019targeted}, and (ii) the multi-objective optimization attack (MOOA) \cite{khare2018adversarial}. All these algorithms are often used in targeted attack scenarios as discussed in \cite{esmaeilpour2020class}.

\section{Proposed Distortion Condition and Adversarial Attack}  
\label{sec:proposed}
This section introduces an extension for the distortion condition of the adversarial attack formulation (Eq.~\ref{eq:generalformulation}) for end-to-end speech-to-text systems in targeted and non-targeted scenarios. This condition fits well for the optimization formulation of the white-box adversarial attack scenario. Our motivation for developing such a distortion condition is threefold: improving the robustness of the adversarial speech signals after playbacks over-the-air, avoiding costly EOT operations, and keeping the quality of the crafted adversarial signal as close as possible to the ground truth input signals. Toward this end, we firstly introduce an integral probability metric (IPM) to measure discrepancies between the adversarial and original signals. Then, we build our distortion condition for adversarial attacks based on this IPM. We explain all the required details in the following subsections.

\subsection{Cram\'{e}r Integral Probability Metric (Cram\'{e}r-IPM)}
One of the standard statistical approaches in measuring the dissimilarity between two probability distributions regardless of the total number of their independent variables is using an IPM \cite{muller1997integral,dodge2006oxford}. Formally, an IPM is a measure for approximating the discrepancies between two (generalizable to higher orders) probability density functions $\mathbb{P}(\cdot)$ and $\mathbb{Q}(\cdot)$ as \cite{muller1997integral,dedecker2007empirical}:
\begin{equation}
    \sup_{f\in \mathcal{F}} \left [ \mathbb{E}_{\vec{x}_{i} \sim \mathbb{P}} f(\vec{x}_{i})- \mathbb{E}_{\vec{x}_{i} \sim \mathbb{Q}} f(\vec{x}_{i}) \right ]
    \label{eq:IPM}
\end{equation}
\noindent where $f(\cdot)$ is called the critic function and it analytically compares the dissimilarity between $\mathbb{P}(\cdot)$ and $\mathbb{Q}(\cdot)$. Moreover, $\mathcal{F}$ denotes the possible function class for the critic and it is completely independent to both the abovementioned probability distributions \cite{sriperumbudur2012empirical}. Mathematically, there are many choices for the function class, however we opt to Cram\'{e}r ($\mathcal{F}_{Cr}$) due to its simplicity, differentiability, and generalizability \cite{szekely2003statistics,bellemare2017cramer}. The statistical definition for $\mathcal{F}_{Cr}$ in the closed form is as \cite{bellemare2017cramer,rizzo2016energy,cramer1928composition}:
\begin{equation}
    \mathcal{F}_{Cr}=\left \{ f_{\vartheta}:\mathcal{X}\rightarrow \mathbb{R}, \mathbb{E}_{\vec{x}_{i}\sim \mathbb{P}}\left ( D^{(1)}f_{\vartheta}(\vec{x}_{i})\leq 1 \right ) \right \}
    \label{eq:cramer}
\end{equation}
\noindent where $D^{(1)}$ indicates the first-order derivation operator and the critic function $f_{\vartheta}$ is smooth with the zero boundary condition \cite{szekely2013energy}. Moreover, $\vec{x}_{i} \in \mathbb{R}^{n \times m}$ is an $m$-channel signal with the length $n$ and $\mathcal{X}$ is a compact subset in $\mathbb{R}$. According to this definition, $\mathcal{F}_{Cr}$ restricts the derivative of $f_{\vartheta}(\cdot)$ within a unit ball to enforce its continuity for higher degrees of $\vartheta$ \cite{bellemare2017cramer,cramer1928composition}. 

Assuming the probability distribution functions for the original and adversarial signals are represented by $\mathbb{P}(\cdot)$ and $\mathbb{Q}(\cdot)$. Therefore, minimizing over Eq.~\ref{eq:IPM} using the $\mathcal{F}_{Cr}$ reduces dissimilarities between random pairs of $\vec{x}_{\mathrm{adv}}$ and $\vec{x}_{\mathrm{org}}$. However, this minimization procedure's convergence is highly dependent on the availability of $f_{\vartheta}$. One possible approach for finding this critic function could be training a neural network (mainly in the generative model frameworks \cite{bellemare2017cramer,salimans2018improving}), Nevertheless, it imposes unnecessary complications and computational overhead to the adversarial optimization formulation. To tackle this issue, we empirically approximate $f_{\vartheta}$ with the joint cumulative distribution function (CDF \cite{deisenroth2020mathematics}) of $\mathbb{P}(\cdot)$ and $\mathbb{Q}(\cdot)$ as the following.
\begin{equation}
    f_{\mathbb{PQ}}(\cdot)\simeq \sum_{i=1}^{n_{t}} \mathbb{P}(\vec{x}_{i,org}) + \mu \cdotp \mathbb{Q}(\vec{x}_{c}), \quad \mu \sim \mathcal{U}[-1,1] 
    \label{eq:CDF}
\end{equation}
\noindent where $\vec{x}_{c}$ is a candidate for the adversarial signal $\vec{x}_{\mathrm{adv}}$ achieved through optimizing for Eq.~\ref{eq:generalformulation} and eventually $\vec{x}_{c} \overset{\mu}{\longrightarrow} \vec{x}_{\mathrm{adv}}$. Furthermore, $n_{t}$ refers to the total number of original samples and $\mu$ is a uniform scaling probability prior to avoid dominating $\mathbb{P}(\vec{x}_{i,org}), \forall i$ over $\mathbb{Q}(\vec{x}_{c})$. Using the critic function $f_{\mathbb{PQ}}(\cdot)$ in Eq.~\ref{eq:cramer} provides a meaningful space for measuring discrepancies between original and adversarial distributions (see similar note in \cite{mroueh2017sobolev}). Thus, minimizing over Eq.~\ref{eq:IPM} maps $\vec{x}_{c}$ onto the original signal manifold and yield a more robust adversarial signal (We discuss this claim in Section~\ref{sec:experiments}).

\subsection{Distortion Condition Using the Cram\'{e}r-IPM}
In this subsection, we introduce our distortion condition based on the Cram\'{e}r-IPM with the critic function $f_{\mathbb{PQ}}(\cdotp)$. In fact, we extend the relative constraint mentioned in Eq.~\ref{eq:generalformulation} to:
\begin{equation}
     \min_{\delta,f_{\mathbb{PQ}}\in \mathcal{F}_{Cr}}\left \|  \mathbb{E}_{\vec{x}_{i} \sim \mathbb{P}} f_{\mathbb{PQ}}(\vec{x}_{i})- \mathbb{E}_{\vec{x}_{c} \sim \mathbb{Q}} f_{\mathbb{PQ}}(\vec{x}_{c}) \right \|
    \label{eq:fnalDistortionCond}
\end{equation}
\noindent where $l_{\text{dB}}(\vec{x}_{c}) < \epsilon$ and $\vec{x}_{\mathrm{adv}} = \arg \min \vec{x}_{c}$. The intuition behind exploiting this condition is finding the most possibly optimal signal $\vec{x}_{c}$ which not only sounds similar to $\vec{x}_{\mathrm{org}}$ according to the loudness metric $l_{\text{dB}}(\cdot)$, but also lies closer to the original signal manifold. Since $\mathbb{E}_{\vec{x}_{i} \sim \mathbb{P}} f_{\mathbb{PQ}}(\vec{x}_{i})$ incorporates the CDF of original and adversarial signals containing background and room noises, it implicitly learns the impulse responses available in the speech dataset. This also possibly makes bypassing $\delta$ very challenging after playbacks over-the-air. 

From a statistical point-of-view, the proposed distortion condition forces an attack optimization formulation to craft an adversarial signal marginally close to the original signals' distribution. This is for counteracting with adversarial defense algorithms, which measure the distance between distribution manifolds to detect an adversarial signal \cite{esmaeilpour2020class}. These defense approaches are inspired by Ma {\it et al.}~\cite{ma2018characterizing}, where it proves the subspace of adversarial signals is distinct from original and noisy samples \cite{esmaeilpour2020detection}. In other words, it is possible to measure the distance between subspaces using metrics defined in orthogonal decomposition forms (e.g., chordal distance in Schur decomposition space \cite{esmaeilpour2020detection}.) Based on this finding, variants of defense algorithms have been developed and they have shown a great performance against strong white, and black-box adversarial attacks \cite{esmaeilpour2020class}. Therefore, incorporating our proposed distortion condition into the attack optimization formulation (i.e., Eq.~\ref{eq:generalformulation}) helps to yield a more robust adversarial signal.

The general overview of our proposed attack algorithm is shown in Algorithm~\ref{algorithm:attack1}. Regarding this pseudocode, we do not employ any EOT operations in our optimization formulation since Eq.~\ref{eq:fnalDistortionCond} implicitly captures local and global distributions of the signals available in the comprehensive speech datasets.
\begin{algorithm}[t]
     \caption{Robust adversarial attack with distortion condition using the Cram\'{e}r-IPM. 
     }
     \begin{algorithmic}[1]
     \small
      \Require {$\vec{x}_{\mathrm{org}}$, $\mathbf{y}$, $\hat{\mathbf{y}}$, $\epsilon$ \Comment{input signal, original phrase, incorrect target phrase, hearing threshold}}
       \Ensure{$\vec{x}_{\mathrm{adv}}$ \Comment{adversarial speech signal}}
       \State $\vec{x}_{c} \leftarrow \vec{x}_{\mathrm{org}}$ \Comment{initializing}
       \State initialize $\mu$ \Comment{random latent variable}
       \While {$\hat{\mathbf{y}}= \mathbf{y}$}\Comment{the goal is reaching to $\hat{\mathbf{y}} \neq  \mathbf{y}$} 
       \State $\delta \leftarrow \min_{\delta} \left \| \delta \right \|_{F} + \sum_{i}c_{i}\mathcal{L}_{i}(\vec{x}_{\mathrm{adv}},\hat{\mathbf{y}}_{i})$
       \State $\vec{x}_{c} \leftarrow \vec{x}_{c} + \delta$ \Comment{candidate adversarial signal}
       \While {$l_{\text{dB}}(\vec{x}_{c})> \epsilon$} \Comment{up to reach $l_{\text{dB}}(\vec{x}_{c})< \epsilon$}
       \State draw a random $\mu \sim \mathcal{U}[-1,1]$
       \State $\delta \leftarrow \min_{\delta,f_{\mathbb{PQ}}}\left \|  \mathbb{E}_{\vec{x}_{i} \sim \mathbb{P}} f_{\mathbb{PQ}}(\vec{x}_{i})- \mathbb{E}_{\vec{x}_{c} \sim \mathbb{Q}} f_{\mathbb{PQ}}(\vec{x}_{c}) \right \|$
       \State $\vec{x}_{c} \leftarrow \vec{x}_{c} + \delta$ \Comment{update the candidate signal}
       \EndWhile
       \EndWhile
       \State $\vec{x}_{\mathrm{adv}}\leftarrow \vec{x}_{c}$ \Comment{crafted adversarial signal}

     \end{algorithmic}
     \label{algorithm:attack1}
\end{algorithm}

\begin{table*}[h]
\centering
\caption{Performance comparison of the adversarial algorithms for attacking DeepSpeech, Kaldi, and Lingvo speech-to-text models. Values shown for every metric are averaged over 10 experiments with different $\hat{\mathbf{y}}_{i}$. Targeted and non-targeted attacks are represented by T and NT, respectively. Additionally, the EOT-based algorithms are check-marked. Herein, $n_{ota}$ stands for the total rounds of robustness against consecutive over-the-air playbacks. Outperforming results are shown in bold.}
\begin{tabular}{c||c||c|c|c|c|c|c|c|c}
\hline
      Model                      & Attack          & WER (\%)            & SLA (\%)            & segSNR  & STOI   & LLR    & Type  & EOT          & $n_{ota}$ \\ \hline \hline
\multirow{6}{*}{DeepSpeech} & C\&W \cite{carlini2018audio}           & $78.94\pm 2.01$  & $30.74 \pm 3.16$ & $\mathbf{21.34}$ & $0.86$ & $0.35$ & T     & $-$          & $0$       \\ \cline{2-10} 
                            & Yakura's attack \cite{yakura2018robust} & $80.28 \pm 3.14$ & $35.49 \pm 0.28$ & $19.57$ & $0.82$ & $0.38$ & T     & $\checkmark$ & $3$       \\ \cline{2-10} 
                            & Metamorph  \cite{chen2020metamorph}      & $72.48 \pm 1.06$ & $45.84 \pm 4.71$ & $17.66$ & $0.84$ & $0.36$ & T     & $\checkmark$ & $1$       \\ \cline{2-10} 
                            & GAA \cite{taori2019targeted}             & $65.80 \pm 2.55$ & $48.35 \pm 3.38$ & $17.02$ & $0.79$ & $0.31$ & T     & $-$          & $1$       \\ \cline{2-10} 
                            & MOOA  \cite{khare2018adversarial}           & $68.06 \pm 2.71$ & $47.01 \pm 1.42$ & $18.46$ & $0.81$ & $0.42$ & T/NT & $-$          & $1$       \\ \cline{2-10} 
                            & \bf Proposed        & $\mathbf{88.19 \pm 3.15}$ & $\mathbf{21.69 \pm 3.09}$ & $18.88$ & $\mathbf{0.88}$ & $\mathbf{0.29}$ & T/NT & $-$          & $\mathbf{4}$       \\ \hline \hline
\multirow{2}{*}{Kaldi}      & Imperio \cite{schonherr2020imperio}        & $69.34 \pm 0.47$ & $31.49 \pm 1.36$ & $\mathbf{24.71}$ & $0.91$ & $0.28$ & T     & $\checkmark$ & $2$       \\ \cline{2-10} 
                            & \bf Proposed        & $\mathbf{83.51 \pm 1.44}$ & $\mathbf{25.86 \pm 1.94}$ & $23.16$ & $\mathbf{0.93}$ & $\mathbf{0.27}$ & T/NT & $-$          & $\mathbf{3}$       \\ \hline \hline
\multirow{2}{*}{Lingvo}     & Robust Attack \cite{qin2019imperceptible}  & $84.37 \pm 2.07$ & $28.21 \pm 2.31$ & $19.44$ & $\mathbf{0.85}$ & $\mathbf{0.41}$ & T     & $\checkmark$ & $3$       \\ \cline{2-10} 
                            & \bf Proposed        & $\mathbf{89.73 \pm 1.75}$ & $\mathbf{22.78 \pm 2.62}$ & $\mathbf{21.58}$ & $0.82$ & $0.43$ & T/NT & $-$          & $\mathbf{5}$       \\ \hline
\end{tabular}
\label{table:comparisonAtt}
\end{table*}

\section{Experiments}  
\label{sec:experiments}
This section discusses the performance of our proposed adversarial attack algorithm, which employs the extended distortion condition using the Cram\'{e}r-IPM. We implement Algorithm~\ref{algorithm:attack1} to attack DeepSpeech (Mozilla's implementation), Kaldi, and Lingvo speech-to-text models without using neither RIR nor CIR filter sets.

Although the proposed algorithm resembles a targeted adversarial attack and requires defining an incorrect target phrase ($\hat{\mathbf{y}}_{i}$), it is generalizable to the non-targeted scenario with the assumption of choosing a random phrase for $\hat{\mathbf{y}}_{i}$ other than the ground-truth ($\mathbf{y}_{i}$). Regarding the common practice in the evaluation of adversarial attack developments that craft adversarial signals only for a portion of the given speech datasets \cite{carlini2018audio,qin2019imperceptible,schonherr2020imperio,chen2020metamorph,esmaeilpour2020class}, we also randomly select 1000 samples from Mozilla common voice (MCV \cite{MozillaCommonVoiceDataset}) and LibriSpeech \cite{panayotov2015librispeech} to evaluate the performance of our proposed attack. These two datasets are comprehensive collections containing utterances from different genders, accents, and ages in short and long speech recordings. We equally assign ten incorrect targeted and non-targeted phrases ($\hat{\mathbf{y}}_{i}$) toward crafting $\vec{x}_{\mathrm{adv}}$, for every selected signal $\vec{x}_{\mathrm{org}}$, 

Since the implementations of the benchmarking speech-to-text models are different, thus for attacking these systems we use the CTC loss function ($\mathcal{L}(\cdot)$) for DeepSpeech, and the cross-entropy loss with masking threshold ($\ell_{net} (\cdot)$, $\ell_{m} (\cdot)$) for the Lingvo and Kaldi systems as explained in \cite{qin2019imperceptible,schonherr2020imperio}. The rest of the settings such as defining $\epsilon$ and beam search decoding for output phrases (both $\mathbf{y}_{i}$ and $\hat{\mathbf{y}}_{i}$) follow the instructions explained in \cite{carlini2018audio}. We make the same assumptions in all experiments for a fair comparison to the Robust Attack, Yakura's attack, Imperio, GAA, MOOA, and Metamorph. We implement all the attack algorithms on two machines with four NVIDIA GTX-1080-Ti and two 64-bit Intel Core-i7-7700 (3.6 GHz, Gen.~10) processors with 8$\times$11 GB and 2$\times$64 GB memory, respectively.

We compare the adversarial attack algorithms' performance from two points of view: (i) attack success rate and (ii) adversarial signal quality. For addressing the first view, we measure the word error rate (WER) and sentence level accuracy (SLA) metrics as they have been characterized for such an aim \cite{qin2019imperceptible,derczynski2013twitter}:
\begin{equation}
\begin{aligned}
    \mathrm{WER} = \frac{(D+I+S)}{N}\times 100 \\ \mathrm{SLA}=\frac{n_{c}}{n_{tot}}\times 100
    \label{eq:wersla}
\end{aligned}
\end{equation}
\noindent where the total number of deletions, insertions, substitutions, and reference phrases have been represented by $D$, $I$, $S$, and $N$, respectively. Moreover, $n_{c}$ denotes the total number of adversarial signals which they could successfully attain the same predefined phrase $\hat{\mathbf{y}}$ after passing through the speech-to-text model.

For addressing the second view, we use three quality metrics: segmental signal to noise ratio (segSNR) \cite{baby2019sergan}, short-term objective intelligibility (STOI) \cite{taal2011algorithm}, and log-likelihood ratio (LLR) \cite{baby2019sergan}. The first two metrics compute the absolute quality of the crafted adversarial signals relative to the available ground-truth speeches ($\vec{x}_{\mathrm{org}}$). The main motivation behind using these two objective metrics is a realistic measurement of adversarial signal quality since not necessarily a robust adversarial attack yields a noise-free speech sample. In other words, the crafted $\vec{x}_{\mathrm{adv}}$ should naturally sound like $\vec{x}_{\mathrm{org}}$, which might carry environmental, echo, and hissing noises. Therefore, higher values for segSNR and STOI metrics interpret as the closer quality of $\vec{x}_{\mathrm{adv}}$ to the original signals.

Since these two metrics are not necessarily bounded, comparing adversarial signals' quality may not be tangible enough. We use the LLR, which is scaled between zero and one, in response to this potential concern. There is an inverse relationship between the magnitude of this metric and the quality of the signals. In other words, for adversarial signals close to their associated $\vec{x}_{\mathrm{org}}$, the LLR is fairly low.

Table~\ref{table:comparisonAtt} summarizes our achieved results. As shown in this table, our proposed attack algorithm outperforms the other algorithms in terms of WER and SLA. However, it partially fails against C\&W, Imperio, and the Robust Attack in terms of quality of the crafted adversarial signals. Table~\ref{table:comparisonAtt} also demonstrates that the proposed attack algorithm's robustness is higher than others after multiple consecutive playbacks over-the-air.

\section{Conclusion}
This paper introduced a new adversarial algorithm for effectively attacking the cutting-edge DeepSpeech, Kaldi, and Lingvo speech-to-text systems. Our proposed approach incorporates a novel extension for the relative constraint of the adversarial optimization formulation to improve the crafted signals' robustness after multiple playbacks over-the-air. This extension minimizes over the Cram\'{e}r integral probability metric between the probability distributions of the original and adversarial signals. This minimization operation projects a candidate adversarial signal onto the original speech recordings' subspace to counteract with potential defense approaches that measure the distance between subspaces.

We experimentally demonstrated that the proposed white-box attack algorithm outperforms other advanced algorithms in terms of attack success rate according to WER and SLA metrics. Moreover, the crafted adversarial signals' average quality via our proposed attack is competitive to other algorithms using objective quality metrics of segSNR, STOI, and LLR.

Our approach is EOT-free, and it has shown considerably higher robustness against consecutive playbacks over-the-air compared to other costly EOT-based adversarial algorithms. However, we could not achieve more than four playbacks averaged over the three victim models. We are determined to address this issue in our future works with developing more constraints on the critic function of the Cram\'{e}r function class.



\bibliographystyle{IEEEtran}
\balance
\bibliography{mybib}


\end{document}